\documentclass[%
 aip,pof,
 amsmath,amssymb,
 reprint,%
]{revtex4-1}

\usepackage{graphicx}
\usepackage{dcolumn}
\usepackage{bm}

\usepackage[utf8]{inputenc}
\usepackage[T1]{fontenc}
\usepackage{mathptmx}
\usepackage{amsmath}
\usepackage{physics}
\usepackage{color}
\usepackage{todonotes}
\usepackage{soul}

\begin{document}

\preprint{AIP/123-QED}






\title[Memory effects in a gas of viscoelastic particles]{Memory effects in a gas of viscoelastic particles}





\author{E. Momp\'o}
\affiliation{ 
Gregorio Mill\'an Institute of Fluid Dynamics, Nanoscience and Industrial Mathematics,
Department of Mathematics, Universidad Carlos III de Madrid, 28911 Legan\'es, Spain
}%

\author{M. A. L\'opez Casta\~no}%
\affiliation{%
Departamento de F\'isica
and Instituto de Computaci\'on Cient\'ifica Avanzada (ICCAEx), Universidad de Extremadura, 06071 Badajoz, Spain
}%

\author{A. Lasanta}
\altaffiliation[Also at ]{Gregorio Mill\'an Institute of Fluid Dynamics, Nanoscience and Industrial Mathematics,
Department of Mathematics, Universidad Carlos III de Madrid, 28911 Legan\'es, Spain}
\affiliation{%
Departamento de \'Algebra. Facultad de Educaci\'on, Econom\'ia y Tecnolog\'ia de Ceuta, Universidad de Granada, Cortadura del Valle, s/n. E-51001 Ceuta, Spain
}%

\author{F. Vega Reyes}
\affiliation{%
Departamento de F\'isica
and Instituto de Computaci\'on Cient\'ifica Avanzada (ICCAEx), Universidad de Extremadura, 06071 Badajoz, Spain
}%

\author{A. Torrente}
 \email{etorrent@est-econ.uc3m.es.}
\affiliation{ 
Gregorio Mill\'an Institute of Fluid Dynamics, Nanoscience and Industrial Mathematics,
Department of Mathematics, Universidad Carlos III de Madrid, 28911 Legan\'es, Spain
}%

\date{\today}

\begin{abstract}
We study a granular gas of viscoelastic particles (kinetic energy loss upon collision
is a function of the particles' relative velocities at impact) subject to a stochastic 
thermostat. We show that the system
displays anomalous cooling and heating rates during thermal relaxation processes,
this causing the emergence of thermal memory. In particular, a significant
\textit{Mpemba effect} is present; i.e., an initially hotter/cooler granular gas can
cool down/heat up faster than an in comparison cooler/hotter granular gas.  Moreover,
a \textit{Kovacs effect} is also observed; i.e., a non-monotonic relaxation of the 
granular temperature --if the gas undergoes certain sudden temperature changes
before fixing its value. Our results show that both memory effects have distinct
features, very different and eventually opposed to those reported in theory for
granular fluids under simpler collisional models.
We study our system via three independent methods: approximate solution of the 
kinetic equation time evolution and computer simulations (both molecular dynamics 
simulations and  Direct Simulation Monte Carlo method), finding good agreement 
between them. 
\end{abstract}


\maketitle 


\section{\label{intro}Introduction}

Recently, observations in a number of systems have increasingly focused  
attention on two --apparently-- paradoxical and different memory effects, namely 
the  Mpemba \cite{MO69} and Kovacs  \cite{KAHR79}  effects, which can be referred 
to as \textit{thermal memory effects} \cite{VL19}. 
The memory of a system can be described as its ''ability to encode, access and 
erase signatures of past history" in its current state, which is intrinsically 
related to a far-from-equilibrium behavior \cite{KPZSN19}. 
In this work we study the existence of both the Mpemba effect (ME) \cite{LVPS17} and the Kovacs
effect (KE) \cite{LVPS19} in low density granular fluids (granular gases). A granular gas is
comprised of a large number of particles with typical sizes larger than
$1~\mu\mathrm{m}$. These macroscopic particles undergo  inelastic collisions, so that
mechanical energy is not conserved. Moreover, in the case of a low density gas,
collisions are approximately instantaneous \cite{G03}.


The ME has been characterized in granular gases as follows: an initially hotter
sample can eventually cool faster than an initially comparatively warm one, when both
are put in contact with a granular temperature source that is cooler than both
samples. The ME is typically triggered only for a given set of far-from-equilibrium
initial conditions \cite{LVPS17}. 

Traditional culture has been aware, since long ago, of the 
ME in water undergoing freezing, although it was systematically studied in a scientific
work only more recently. In particular, the effect owes its name to high-school
student Erasto B. Mpemba, who noticed and measured the effect in milk while making
ice-cream in a high-school lab. A complete account of the experimental observations
(also with water experiments) was later published by Erasto B. Mpemba in
collaboration with Prof. Osborne \cite{MO69}. A recent study reveals the
reproducibility of such experiments \cite{BH20}. And yet the explanation of the
underlying physical mechanism triggering the effect has remained elusive for long.  

However, rather recently, Lu and Raz showed that the ME should always be present in
Markovian processes \cite{LR17}. In fact, they also proved that there exists also an
inverse ME; i.e., a sample of a system that is colder than another, otherwise
identical sample may heat up
faster when both are put in contact with the same hotter temperature source. The
direct and inverse ME were almost simultaneously (and independently) confirmed by other authors
in a granular gas of hard particles \cite{LVPS17}. Other systems like carbon nanotube
resonators \cite{GLCG11} or  clathrate hydrates \cite{AKKL16} 
had also shown this interesting effect shortly
earlier, although only in its classical direct variant.

These works have raised renewed interest in this long-time problem; the
ME has been detected in  theoretical models such as non-Markovian 
mean-field systems \cite{Yang20}, driven granular gases \cite{BPRR20,GKG20}, 
molecular gases \cite{SP20}, inertial suspensions \cite{THS21}, 
antiferromagnetic models \cite{Klich19, GR20}, quantum spin models \cite{Nava19} 
and spin glasses \cite{Bal19}, liquid water (not involving a phase
transition) \cite{GLH19} and experiments with colloids experiments \cite{KB20}. In addition, 
the inverse ME has been observed for the first time in experiments \cite{KCB21}, in particular 
in a colloidal system. Interestingly and very recently, it has been proven that classical 
\cite{GKG20} and quantum \cite{CLL21} relaxation dynamics in many particle systems can be 
accelerated using Mpemba-like strategies, opening the door to future direct applications. 

More recently, and in other fields, related memory effects are being actively 
investigated. The existence of the so-called mixed ME, a process where two identical 
systems with initial temperatures respectively above and below their common relaxation 
temperature have relaxation curves that cross each other, has been latterly detected \cite{THS21, GG20}. In the same 
line, asymmetry in thermal relaxation for equidistant temperature quenches has been found 
for systems near stable minima \cite{LG20} and the fact that uphill relaxation (warming) 
is faster than downhill one (cooling) is being currently studied \cite{VvH21,M21}.

Granular dynamics is perhaps the field where recent
literature on the ME is most abundant. Indeed, the ME in granular fluids has also
been analyzed in recently published works on slightly different systems, such as in
confined granular layers \cite{MSB19}, granular gases with rotational degrees of freedom
\cite{Tal19}, granular suspensions \cite{THS21} and particles subject to a
drag force \cite{SP20}. These works provide insight into the same
phenomenon; namely the faster cooling/heating of a comparatively hotter/cooler
system. This results in two (granular) temperature relaxation curves that cross each other
at some point in their evolution, as essentially predicted by the original
work in granular gases \cite{LVPS17}, with the only variation of a subtly different
collisional model or another extra modification, such as boundary conditions (for the
confined layer) or volume forces (as in the case of suspensions and particles subject
to a drag force).

And yet, experimental verification of the ME in granular dynamics
is still pending, unlike in the case of colloids, where experiments have clearly
demonstrated the existence of the ME \cite{KB20}. 
Moreover, this recent experimental observation in colloids, combined
with advances from theoretical works predicting the ME in granular
fluids, encourage conducting laboratory experiments of granular dynamics, in search
of more experimental confirmation.

However, prior to searching for corroboration of the ME in granular
matter, a significant improvement in the collisional model used for the theoretical
detection is much needed. In effect, the collisional models
used so far to describe the ME in granular fluids 
 do not take into consideration the experimental evidence that collisions
usually depend on the particles' relative velocities on impact \cite{W92,FLCA94}. 
These particle models may actually be seen as reductions of the Walton collisional 
model \cite{W92}, which takes into account the velocity dependence only through 
the friction coefficient (which, in turn, depends on the impact angle, for sufficiently 
small grazing angles \cite{GBM15,FLCA94}).

Walton's model is  a simplification of the theory by Maw, Barber, and Fawcett
\cite{MBF81} (although the work by Walton does incorporate the kinetic energy loss
during collisions), which is an extension of Hertz's theory for elastic contacts
\cite{MBF81,D00}. Walton's model includes the effects of oblique impact in detail,
which are usually of relevance at experimental level  grain collisions (an exhaustive 
report of careful measurements of coefficients of
restitution for macroscopic particle collisions may be found in \cite{M9409}).

Although Walton's model makes a fair enough description for the experimental
conditions of collisions in a variety of hard materials (such as metals)
\cite{FLCA94}, it is known that  the normal component of the post-collisional
relative velocity depends significantly, in general, on the impact velocity absolute
value \cite{GBM15}. 
In this sense, Brilliantov, P\"oschel and collaborators worked on collisional models 
for inelastic particles \cite{BSHP96,BP04} where the kinetic energy loss upon 
collision is considered to be dependent on the impact velocity (henceforth, this 
model is referred to as \textit{viscoelastic particle} model) \cite{BP04}.


On a different note, there is also an extensive literature on the KE,
which was originally detected in a polymer system by Andr\'e J. Kovacs and 
co-workers \cite{K63,KAHR79}, in an experiment described as follows. 
A sample of polyvinyl acetate, initially in a thermal equilibrium state (with 
known temperature $T_0$) is subject to a temperature drop, to a value $T_{1}\ll T_0$. 
While the polymer is still relaxing towards the new equilibrium state, the
temperature is suddenly increased, at a (waiting) time $t_{w}$, to an intermediate 
value $T_\mathrm{st}$, with $T_{1}< T_\mathrm{st} < T_{0}$. 
This temperature $T_\mathrm{st}$ is maintained stationary until the polymer 
reaches a final equilibrium state. 
The trick in their set of experiments is that when $T_\mathrm{st}$ is applied 
at time $t_w$, the instantaneous volume $V(t=t_w)$ equals the stationary value 
that the volume should have for $T_\mathrm{st}$; i.e., $V(t=t_w, T_\mathrm{st}) 
= V_\mathrm{st}$. 
In this way, both $T(t=t_w)$ and $V(t=t_w)$ are equal to their stationary value 
and hence, as temperature is kept by means of a temperature source, no further 
evolution for the volume would be expected.  
Instead, the volume follows a non-monotonic time evolution; in this case, what Kovacs
and co-workers observed was that the volume rebounds up to a maximum, decreasing
afterwards towards its final equilibrium value $V_\mathrm{st}=V(t=t_w)$.

This non-monotonic behavior, later denominated \textit{Kovacs hump}, consists 
therefore in the volume reaching \textit{one} maximum before returning to its 
equilibrium value ${V}_\mathrm{st}$. 
As these authors explained, such a behavior is due to the fact that the polymer 
has complex dynamics and there are additional relevant variables (other than 
volume and temperature) involved in the relaxation process, these being coupled 
with each other and with the temperature \cite{KAHR79}.

More recently, the KE has been reported in several other complex systems such as 
glassy systems \cite{MS04, ALN06, BM18, SXHLWEW20}, active matter \cite{KSI17} 
or in the thermalization of the 
center of mass motion of a levitated nanoparticle \cite{MLFBNR21}; thus, one could 
expect this memory effect to appear in other real systems. 
Additionally, an analogous effect has been observed in the temperature time evolution
of  other athermal systems \cite{KSI17, PP17} and granular fluids subject to a sudden
temperature change \cite{PT14, TP14,  Brey14, LVPS19, SsP20}. However, the collisional 
models in these works have neglected the effects of impact velocity on the collision 
inelasticity. 
We will analyze here the KE in
a granular system for the more realistic viscoelastic collisional model, in search of a
more definitive confirmation of this effect in granular dynamics. Furthermore, as we 
will see, both ME and KE pertain to the same class of thermal memory effects.

Therefore, in the next section we describe our system and the theoretical basis 
in more detail. 
Section~\ref{MpembaSection} is devoted to the analysis of our results involving 
the presence of a clear Mpemba-like effect in our viscoelastic granular system, 
where we have considered both a cooling and a heating protocol. 
In Section~\ref{KovacsSection} we investigate the existence of a Kovacs-like 
effect and its relationship with the parameters characterizing the system. 
Finally, Sections~\ref{discussion} and \ref{conclusion} are dedicated to discussion 
and final conclusions.

\section{System and time evolution equations}
\label{system}

We consider a system of thermalized grains, with a very low particle density ($n$) 
at all times. In our system,  all particles are identical spheres of mass $m$ and diameter 
$\sigma$ and, apart from having a mesoscopic size, there is no restriction regarding 
the value of their diameter. With low particle density in this context one usually means 
\cite{G03} that contacts occur only between two particles and contact time is negligible 
as compared to the typical time between collisions. The fact that such collisions 
are instantaneous and binary allows ignoring velocity correlations and thus, 
assuming the \textit{molecular chaos} ansatz \cite{G03}, considering a statistical 
description based on a single particle velocity distribution function $f(v,t)$. 

Since particles are not microscopic, collisions are \textit{inelastic}; i.e., energy 
is not preserved. Furthermore, the degree of inelasticity  in each collision 
experimentally depends on particles' relative velocities at impact, \cite{GBM15} $v_{12}$. 
For this reason, collisions are best described in this case if the effect of 
relative velocity on collision inelasticity is taken into account. In particular, 
a velocity-dependent restitution coefficient, in the model by Brilliantov \& 
P\"oschel is given by \cite{BSHP96, BP04}
\begin{equation}
\epsilon = 1 - C_1 A \alpha^{2/5} | \text{$v_{12}$} \cdot e_{12} | ^{1/5} + C_2 A^2 
\alpha^{4/5} | \text{$v_{12}$} \cdot e_{12} | ^{2/5} \pm \dots,
\end{equation}
where $e_{12}$ is a unit vector in the direction of the colliding particles' relative position vector, $A$ is a material dependent dissipative constant, $\alpha=\left(
\frac{3}{2} \right)^{3/2} \frac{Y \sqrt{\sigma}}{m(1-\nu^2)}$, $Y$ is the 
Young modulus, $\nu$ is the Poisson ratio and  $C_1=1.15344$ and $C_2 = \frac{3}{5} 
C_1^2 = 0.79825	
$ are known constants; the terms containing higher powers of $A$ can be
neglected for small enough $A$. In summary, the restitution coefficient can be 
described as \cite{DBPB13}
\begin{equation}
\label{restitution}
\epsilon \simeq 1 - \gamma_{\nu} | \text{$v_{12}$} \cdot e_{12} | ^{1/5} + (3/5)\gamma_{\nu}^2 |
\text{$v_{12}$} \cdot e_{12} | ^{2/5},
\end{equation}
where the known dissipative coefficient $\gamma_{\nu} = C_1 A \alpha^{2/5}$ 
depends on the material properties. It is however possible, provided that dissipation is not too large\cite{DBPB13}, to reduce the 
velocity-dependent equation \eqref{restitution} to an
approximate expression that depends only on the velocity ensemble average 
through a dependence on the granular temperature $T(t)$ 
\begin{equation}
  \epsilon_\mathrm{eff}= 1+ \sum_{k=1}^{N_\epsilon}B_k\text{$\eta$}^{k/2}\left[2T(t)/T_0\right]^{k/20},
\label{epsilon_eff}
\end{equation} 
with $\eta=\frac{\gamma_\nu}{C_1}\left(\frac{T_0}{m}\right)^{1/10}$ the dissipation coefficient and $T_0 = T(0)$  the initial
temperature of the system. Expression \eqref{epsilon_eff} was deduced by Dubey
\textit{et al.} \cite{DBPB13}, in whose  manuscript the values of the coefficients $B_k$ can 
also be found. We assume that inelasticity is not large so that the approximation in \eqref{epsilon_eff} remains accurate.

Thermalization of grains is achieved by means of the action of a homogeneous stochastic force, $\mathbf{F}^\mathrm{wn}$. 
In this case, the thermostat is modelled as a zero-mean Gaussian white noise \cite{MS00}: 
\begin{equation}
  \label{whitenoise}
 \left< \mathbf{F}^\mathrm{wn} \right> = \mathbf{0}, \quad \left<
   \mathbf{F}_i^\mathrm{wn}(t)\,\mathbf{F}_j^\mathrm{wn}(t') \right> = \delta_{ij}\delta(t-t')
   \xi_0^2\mathbf{I},
\end{equation} 
where $\mathbf{I}$ is the $3\times 3$ unit matrix, $\delta_{ij}$ is the 
Kronecker delta and $\delta(t)$ is the Dirac delta function; $\xi_0$ characterizes the strength of the stochastic force. 

The appropriate  kinetic equation for our low density granular gas is the Boltzmann
equation \cite{BP04}, where there is an additional term that takes into account 
the action of the stochastic thermostat 

\begin{equation}
  \label{kinetic}
\frac{\partial f(v,t)}{\partial t} = J[f,f] +
\frac{\xi_0^2}{2}\frac{\partial^2f(v, t)}{\partial v^2},  
\end{equation} 
in which $f$ is the particle velocity distribution function and 
$J[f,f]$ is the collision integral for viscoelastic spheres \cite{BP04}, with 
collisions modeled in this case by taking into account a coefficient of restitution 
according to \eqref{epsilon_eff}, as we said. We have also considered that the system 
is at constant density and homogeneous at all times, and therefore, in our case, $f(\mathbf{r}, 
\mathbf{v},t) = f(v,t)$.

As is known, the stationary distribution of a granular
gas of viscoelastic particles typically differs \textit{very slightly}
from the Maxwell-Boltzmann
distribution, this difference being noticeable only in the high energy tails \cite{BP04,DBPB13}.
(Only under extreme non-equilibrium conditions do deviations from the Maxwellian 
become more significant). Such deviations can be quantified as follows.
The set of associated Laguerre polynomials (called also \textit{Sonine}
polynomials \cite{BP04,S80}) $L_p^{(k)}(x)$ fulfills
the orthogonality conditions $\int_0^\infty  x^k
e^{-x}L_p^{(k)}(x)L_q^{(k)}(x)\mathrm{d}x=\delta_{p,q}(p+k)!/p!$. 
Therefore, they can be
used to express accurately $f(v)$ (with $x\equiv mv^2/2T$) as a
truncated expansion (usually called Sonine expansion in the context of kinetic
theory of gases \cite{G03}) in the form
\begin{equation}
  \text{$f(c)$} = \phi(c)\left[1+\sum_ {p=1}^\infty a_p S_p(c^2)\right],
  \label{sonine_expansion}
\end{equation} 
where $c$ is the scaled velocity $c\equiv v/v_T$, 
$v_T=\sqrt{2T/m}$ is the thermal temperature  and 
$\phi(c)\equiv n/(v_T\pi^{1/2})^3e^{-c^2}$
is the Maxwell distribution expressed as a function of $c$.
 Also,
$S_p(c^2)$ represents an associated Laguerre polynomial $L_p^{(k)}(c^2)$ of order $k=1/2$. 
Note that in this case $k=1/2$ is the appropriate choice, since it
yields $a_p=0$ for all $p$ if the distribution function is the Maxwellian
\cite{NE98}.

In other words, the coefficients $a_p$, which are denoted as \textit{cumulants},
measure the deviation of $f(v)$
off the Maxwellian. Henceforth, we will deal only with \textit{slightly} non-Maxwellian states, for which 
we need to retain only the first two cumulants; i.e., we use \cite{CVG13}
\begin{equation}
a_2 = \frac{4}{15}\langle c^4\rangle-1, \quad \quad a_3=\frac{4}{5}\langle
c^4\rangle-\frac{8}{105}\langle c^6\rangle -2, \quad \quad  a_{p>3}\simeq 0.
    \label{cumulants}
\end{equation}

The thermal evolution of the system can be described by means of  the differential equations
that result from integration of the moments of the kinetic equation \eqref{kinetic} 
within the degree of approximation in \cite{BP04,NE98}:
\begin{align}
\frac{dT}{dt} &= -\frac{\widehat{\kappa}}{3}\mu_2 T^{\frac{3}{2}} + m\xi_0^2, 
\label{eq:main_ode1}\\
\frac{da_2}{dt} &= \frac{2}{3}\widehat{\kappa}(1+a_2)\mu_2\sqrt{T} - \frac{2}{15}
\widehat{\kappa}\mu_4\sqrt{T} - 2\frac{a_2}{T}m\xi_0^2, \label{eq:main_ode2}\\
\begin{split}
    \frac{da_3}{dt} &= \widehat{\kappa}(1-a_2+a_3)\mu_2\sqrt{T} - \frac{2}{5}
\widehat{\kappa}\mu_4\sqrt{T} \\
& + \frac{4}{105}\widehat{\kappa}\mu_6\sqrt{T} - 
3\frac{a_3}{T}m\xi_0^2,
\end{split}
\label{eq:main_ode3}
\end{align} where $\mu_p$ is the $p$-th moment of the collision integral, 
linearized in \cite{DBPB13} as:
\begin{equation}
\begin{split}
    \mu_p(T,a_2,a_3) =& \sum_{k=0}^{20}\Big(
    	M_k^{(p,0)} + M_k^{(p,2)}a_2 + M_k^{(p,3)}a_3 + M_k^{(p,22)}a_2^2 \\
    	& + M_k^{(p,33)}a_3^2 + M_k^{(p,23)}a_2a_3
    \Big)
    \left[
    	\frac{\gamma_\nu}{C_1}\left(
    		\frac{2 T}{m}
    	\right)^{\frac{1}{10}}
    \right]^{\frac{k}{2}}.    
\end{split}
\label{eq:mu_p}
\end{equation}
Here, $M_k^{(p,j)}$ are known numerical constants \cite{DBPB13} and
we have defined $\widehat{\kappa}\equiv2\sqrt{2}\sigma^2 n /\sqrt{m}$.

When a steady-state is reached ($d/dt=0$), all of the moments of the distribution function
become time independent \cite{VSK14}. We label the steady state parameters with
superscript ``$\mathrm{st}$''. With some little algebra, we obtain
\begin{equation}
  \mu_2^\mathrm{st} = \frac{3m\xi_0^2}{\widehat{\kappa} \left(T^\mathrm{st}
  \right)^{\frac{3}{2}}}, \quad 
 \mu_4^\mathrm{st} = 5\mu^\mathrm{st}_2, \quad  \mu_6^\mathrm{st} =
 \frac{105}{4}(a_2^\mathrm{st}+1)\mu^\mathrm{st}_2,
\label{mupst}
\end{equation} where $\mu_2^\mathrm{st}$ has previously been obtained 
from \eqref{eq:main_ode1} with $dT/dt=0$.

Expressions \eqref{mupst} can be solved numerically with a fixed point iteration scheme 
by means of the approximation given by equation~\eqref{eq:mu_p}.

\subsection{Initial conditions and thermal memory}
\label{subsec:ICM}

From a mathematical point of view, analyzing a thermal relaxation process in our
system  involves solving the differential equations
\eqref{eq:main_ode1}\textendash\eqref{eq:main_ode3}. For this, we logically
need to specify the initial conditions triplet $(T(0),a_2(0),a_3(0))$, after which the 
relaxation process is fully specified  (its relaxation is unique). 

Therefore, the thermal memory emerging from the ME and KE in granular dynamics
strictly does not recall (does not provide information about) past states, i.e., states, at
$t<0$, prior to its final relaxation. This is in contrast with the persistent memory of 
past history in  other systems, such as foam sheets, which can remember past shear cycles and
have the ability, under the appropriate conditions, to mimic them \cite{PHLN20}.

Thus, for a comprehensive study of thermal memory in granular dynamics, we will
explore wide intervals of the initial triplet $(T(0),a_2(0),a_3(0))$. Moreover, we
already know not all initial states can produce memory effects \cite{KPZSN19,LVPS17}. In
fact, ME and KE usually occur during a relaxation process off an initial
state that is sufficiently far from the stationary state (henceforth, 
by \textit{far} in this context we mean an initial temperature that is not close to 
the stationary temperature) to which the system is going to
relax \cite{KAHR79,LVPS17}. This feature can be characterized attending to the
initial values of the relevant distribution moments vs. their final
values. Therefore, our study will pay attention to the relative differences 
$\Lambda = (T^{rel}, a_2^{rel}, a_3^{rel})  =
( |T(0)-T^\mathrm{st}|/T^\mathrm{st},
|a_2(0)-a_2^\mathrm{st}|/a_2^\mathrm{st}, |a_3(0)-a_3^\mathrm{st}|/
a_3^\mathrm{st})$.

\subsection{Initial conditions and experiments/computer simulations}

Analyzing thermal relaxation processes for different ranges of the initial triplet 
$\Lambda$ out of the
differential equations \eqref{eq:main_ode1}\textendash\eqref{eq:main_ode3} is
relatively straightforward since it only involves specifying the numerical values of
$(T(0),a_2(0),a_3(0))$ and $\xi_0^2$, the noise intensity, which sets the final state and
therefore $(T^\mathrm{st}, a_2^\mathrm{st}, a_3^\mathrm{st})$.

Undertaking the analogous task in experiments and computer simulations is,
notwithstanding, somewhat more subtle. In particular, in experiments we can indeed control the
degree of thermalization of the system (here, this role is played by the parameter
$\xi_0^2$). However, in experiments there is no complete control on the initial
microscopic state and thus the triplet $\Lambda$
cannot in general be set at will to specific values, except for the
temperature. All we can do
is to make the system undergo subsequent thermal
\cite{PT14,SR20} or pressure source \cite{K63} changes, so we can produce a far from
equilibrium initial state with \textit{a priori} unknown values of $(T(0),a_2(0),a_3(0))$.

Luckily enough, in particle simulations we can always control the particles
initial velocities as well, and in this way we have control  of the initial
velocity distribution function, but not of its moments. Thus, the most efficient way to produce 
an initial distribution function with the desired values of $(T(0),a_2(0),a_3(0))$ is with the
use of random variate distribution functions \cite{B94}.

Since the stationary absolute values of the cumulants $(a_2^\mathrm{st},
a_3^\mathrm{st} )$ tend to be rather small (close to zero) \cite{DBPB13}, we would need 
distributions that can attain at least moderate values of
$(|a_2(0)|, |a_3(0)|)$, so that the values of  
$(a_2^{rel} , a_3^{rel})$ are not close to zero.
Thus, a good choice is the random variate of a Gamma distribution, since it can achieve such
values for $(a_2(0), a_3(0))$ \cite{MKL}. The Gamma distribution function  with shape parameter $a$, shift parameter $\varsigma$ and scale parameter $\beta$ may be expressed in 
terms of the variable $v^2$  as \cite{MKL}
 \begin{equation}
     f_{a, \varsigma, \beta}(v^2) = 
     \begin{cases}
      \frac{1}{\Gamma(a)\beta^a}(v^2-\varsigma)^{a-1}e^{-(v^2-\varsigma)/\beta},  & \ v^2 > \varsigma \\
       0, &\ v^2 \le \varsigma
     \end{cases}.
 \end{equation} 
The angular parts (in spherical coordinates) of
particle velocities are drawn from random variates of a uniform distribution, since
the system is isotropic. Therefore, for computer simulations in this work, we use the
random variates of a Gamma distribution and of a uniform distribution for generating
the  initial particle velocity modulus and angular part, respectively.


\section{Mpemba Effect in a granular gas of viscoelastic particles}
\label{MpembaSection}

Since the ME occurs during a thermal transient, given the noise
intensity $\xi_0$ and inelasticity $\epsilon$, we need to set
an initial temperature different from its corresponding stationary value. Thus, we will consider two variants of the 
protocol for choosing initial conditions: 
a) a heating process ($T(0)<T^\mathrm{st}$); and 
b) a cooling process ($T(0)>T^\mathrm{st}$).

This is not enough, however, since (as demonstrated in a former work for
a granular fluid composed of hard particles with
constant coefficient of restitution \cite{LVPS17}) the memory effect will emerge only
under particular conditions, among which it is necessary that the shape of
the initial distribution function is far from its steady state form; i.e., the
relative differences of the first distribution function moments, and hence
$(a_2^{rel}, a_3^{rel})$, need to be big enough \cite{a3plays}. 

Therefore, the question remains if an anomalous temperature relaxation rate can be observed when
the granular gas is driven off its steady state distribution function. By anomalous we mean here that the
transient from the initial state with $T(0)^\mathrm{closer}$ will be
longer than the transient resulting from $T(0)^\mathrm{further}$, with
$|T(0)^\mathrm{closer}-T^\mathrm{st}|<|T(0)^\mathrm{further}-T^\mathrm{st}|$. This
implies, for instance, that a hotter system can cool more quickly than a warmer one, 
which is the counter-intuitive effect known from Antiquity in water. 
Conversely, it also implies that a colder system may heat up more quickly than a warmer one.

For each of these two protocol variants (heating and cooling transients), we will
compare the evolution of the same physical system  (in this case, granular gas
with the same inelasticity) evolving off different initial
states. We used there three different sets of initial conditions for each protocol
variant. We compare the three resulting different transients in pairs, in such a way that one of the 
initial states to be compared with has a temperature
$T(0)^\mathrm{closer}$ and the other $T(0)^\mathrm{further}$. The ME is
indistinctly identified as a crossing between such two transients (the \textit{normal} transient
temperature relaxation rate during transients would not produce such a crossing since
the initial state with temperature closest to its stationary value would relax faster).

\begin{figure}[!ht]
\centering 
\includegraphics[width=0.5\textwidth]{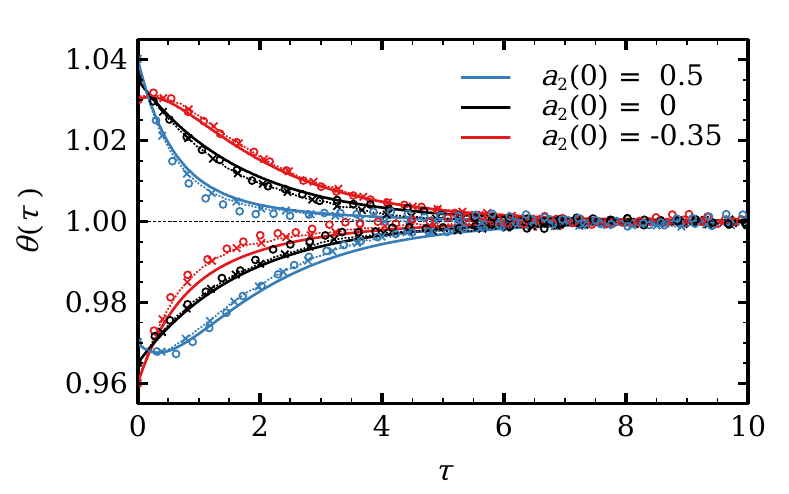}
\caption{Evolution of the scaled temperature $\theta$ towards the steady state 
showing a clear Mpemba-like effect in a cooling process (top curves, with initial
states $(\theta(0), a_2(0),  a_3(0))=(1.04, 0.5, -0.071)$, $(1.035, 0, 0)$ and 
$(1.03, -0.35, -0.375)$) and an inverse Mpemba-like effect in a heating process 
(bottom curves,  with initial states $(\theta(0), a_2(0), a_3(0))=(0.97, 0.5, 
-0.071)$, $(0.965, 0, 0)$ and $(0.96, -0.35, -0.375)$). Solid lines represent 
the exact solution, open circles correspond to MD, and dotted lines with crosses 
refer to DSMC simulations.  }
\label{fig:mpemba}
\end{figure}

In this sense, Figure~\ref{fig:mpemba} depicts the evolution
of the temperature ratio $\theta(\tau)=T(\tau)/T^\mathrm{st}$  for three 
transients for each protocol. Here, time has been expressed as
$\tau\equiv\widehat{\kappa}\sqrt{T^\mathrm{st}}t$, i.e., as number of collisions per
particle, referenced to the steady state, since 
$\nu^{\mathrm{st}}\equiv\widehat{\kappa}\sqrt{T^{\mathrm{st}}}$ is the collision 
frequency for a system at temperature $T^{\mathrm{st}}$. Results are shown for both 
cooling (top curves) and heating (bottom curves) processes as obtained via three 
independent methods: solution of the system of
equations  \eqref{eq:main_ode1}\textendash\eqref{eq:main_ode3},  molecular dynamics
simulations (MD) and exact numerical solution of the kinetic equation obtained from
the Direct Simulation Monte Carlo (DSMC) method (see Appendix for more details on our 
numerical technique and MD and DSMC simulations).


As we can see, our results exhibit an excellent agreement between the three 
independent methods, clearly displaying the ME  in both the cooling and
heating processes (the latter referred to as \textit{inverse ME}). In
effect, curve crossings between transients are clearly observed, in this case at very
early times (before $\tau\sim1$, i.e., before all particles have statistically had 
the chance to collide once after initialization).

For the cooling variant ($T(0)^\mathrm{further} > T(0)^\mathrm{closer} >
T^{\mathrm{st}}$), we have
represented  the triplets $(1.04, 0.5, -0.071)$, $(1.035, 0, 0)$ and $(1.03, -0.35, -0.375)$, whereas
for the heating protocol ($T^{\mathrm{st}} >  T(0)^\mathrm{closer} > T(0)^\mathrm{further}$) we have
used the triplets $(0.96, 0.5, -0.071)$, $(0.965, 0, 0)$ and $(0.97, -0.35,
-0.375)$. Notice that the relaxation curve of the initial state whose temperature is
furthest from the stationary value also crosses the one having
intermediate Maxwellian initial state. This is an interesting result that indicates
that the initial conditions further away from the steady state can also produce an ME
relative to an initial state that resembles a microscopic equilibrium state.

\begin{figure}[!ht]
\centering 
\includegraphics[width=0.5\textwidth]{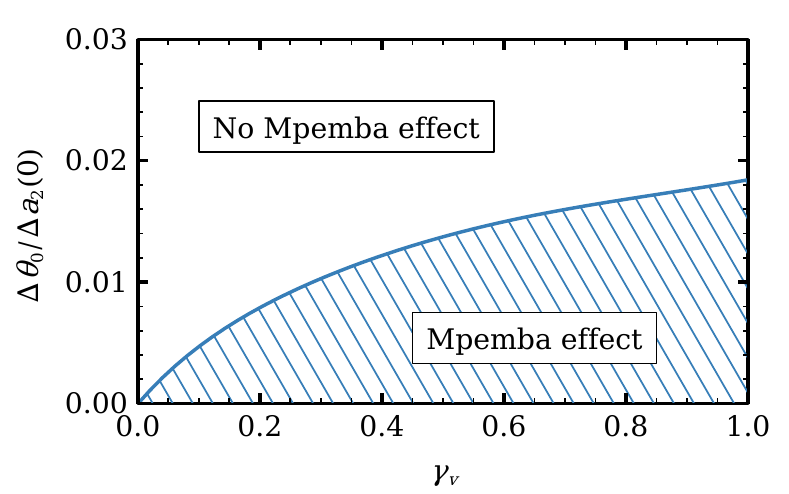}
\caption{Theoretical phase diagram in the plane $\Delta \theta(0) / \Delta a_{2}(0)$ 
vs $\gamma_\nu$. The regions of the plane inside which there appears or does not appear 
the ME are separated by a curve, showing that there is a maximum of the parameter for
which  the ME can be observed.}
\label{fig:mpembaphase}
\end{figure}

We analyze now to what extent the ME is observable in a granular gas of
viscoelastic particles. For this, we determine, from integration of  
equations~\eqref{eq:main_ode1}-\eqref{eq:main_ode3}, the ratio 
$\Delta\theta(0)/\Delta a_2(0)$, where
$\Delta\theta(0)\equiv (T_A(0)-T_B(0))/T^\mathrm{st}$ and $\Delta a_2(0)\equiv
a_{2A}(0)-a_{2B}(0)$ as a function of the dissipative coefficient
$\gamma_\nu$. Results are displayed in Figure \ref{fig:mpembaphase}, where we can see
there is a wide region of the parameter space in which the ME is
present. Much like in the case of a gas of hard spheres, the Mpemba region grows
as inelasticity increases ($\gamma_\nu=0$ is the perfectly elastic
collision limit, and inelasticity grows as $\gamma_\nu$) but here, distinctively,
the width of the Mpemba region for higher inelasticities soon
reaches an asymptotic value and thus increasing inelasticity stops yielding wider
Mpemba regions. Moreover, the ME  is in general smaller for viscoelastic
particles ($\Delta\theta(0)/\Delta a_2(0)<0.02$, even for higher inelasticities)
which is less than $25\%$ the size of the ME that has been detected at
high inelasticities in granular gases of hard spheres\cite{LVPS17}.


\section{Kovacs effect in a granular gas of viscoelastic particles} \label{KovacsSection}

As we already know, the behavior of the KE is, in general, rather more
complex in granular fluids than that of the original KE in a polymer layer
\cite{K63,KAHR79}. For instance, granular fluids display an upwards hump, analogously
to the behavior encountered in the original work by Kovacs (and thus called
henceforth \textit{normal hump}); but also an \textit{anomalous hump} downwards was
found, this consisting in further cooling after the bath temperature is fixed during a
cooling transient in systems with high inelasticities
\cite{PT14,TP14}. Furthermore, multiple --alternatively upwards and
downwards-- humps have been detected in granular fluids where particles have
rotational degrees of freedom as well \cite{LVPS19}. But, again, the question remains if the
KE is really present in granular fluid experiments and, as we said, a first
theoretical approach is to model grain collisions more realistically, via the
viscoelastic model in this case.


\begin{figure}[!htb]
\centering 
\begin{tabular}{l}
(a)\\ 
\includegraphics[width=0.49\textwidth]{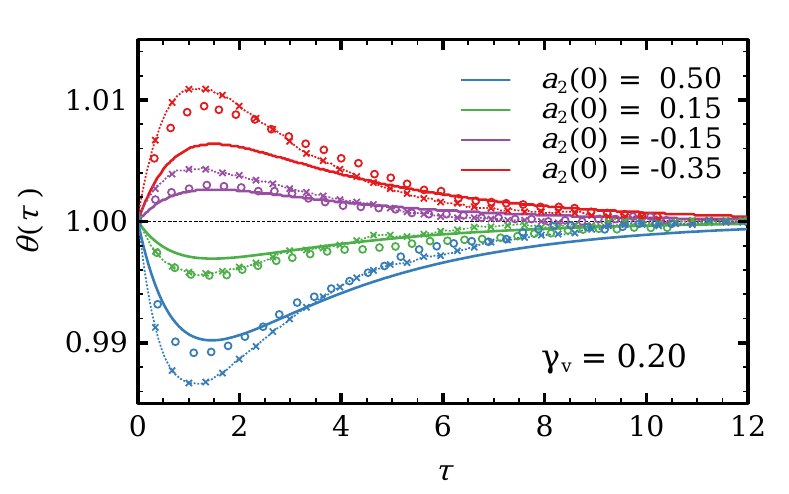}\\
(b)\\
\includegraphics[width=0.49\textwidth]{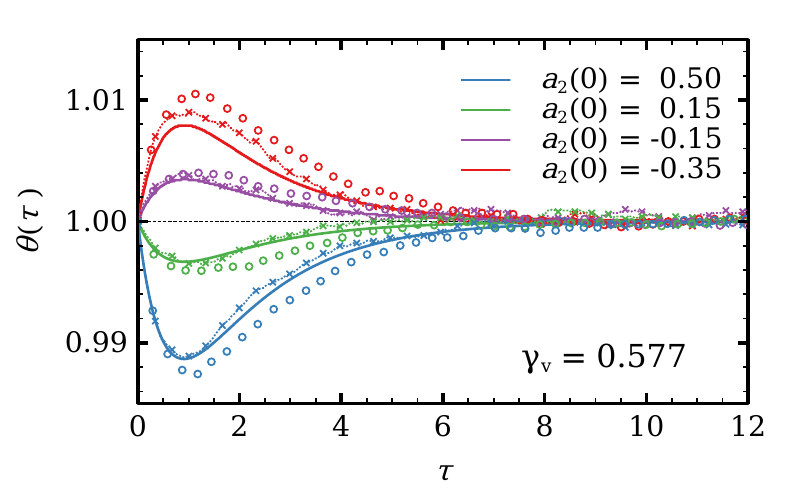}\\
\end{tabular}
\caption{
Theoretical (lines),  MD simulation (open circles) and DSMC (dotted lines with crosses) 
results, showing the KE for (a) $\gamma_\nu=0.2$ and (b) $\gamma_\nu=0.577$.} 
\label{fig:kovacs}
\end{figure}

Thus, in order to detect the KE in granular fluids with the (more realistic)
viscoelastic model, we subject the system to a
stochastic thermostat of intensity $\xi_0^2$, which is set so that
$T(0)=T^\mathrm{st}(\xi_0^2)$. In addition, the initial values of the cumulants,
$a_2(0)$ and $a_3(0)$, are set to values very different from 
their steady state values (similarly to the ME
case), so that the initial distribution function is far from its stationary
form. Therefore, once the system is left to evolve, it will undergo a transient with
the necessary time for the cumulants to approach their steady state values (which are
known \textit{a priori} since they only depend on the dissipation coefficient \cite{DBPB13}).

But, since the system is already initially at its steady state temperature, any
eventual departure during the transient of the granular temperature at later times
should correspond to a Kovacs-like effect. This is potentially possible due to the
strong coupling of the time evolution of the cumulants $a_2$ and $a_3$ and $T$ (see
Equations~\eqref{eq:main_ode1}-\eqref{eq:main_ode3}). And, as we will show, we have
certainly observed the KE for wide ranges of the parameter space for viscoelastic 
particles.


\begin{figure}[t!]
\begin{tabular}{ll}
(a) \\ 
\includegraphics[width=0.48\textwidth, trim= 3.5mm 0 1.5mm 0, clip]{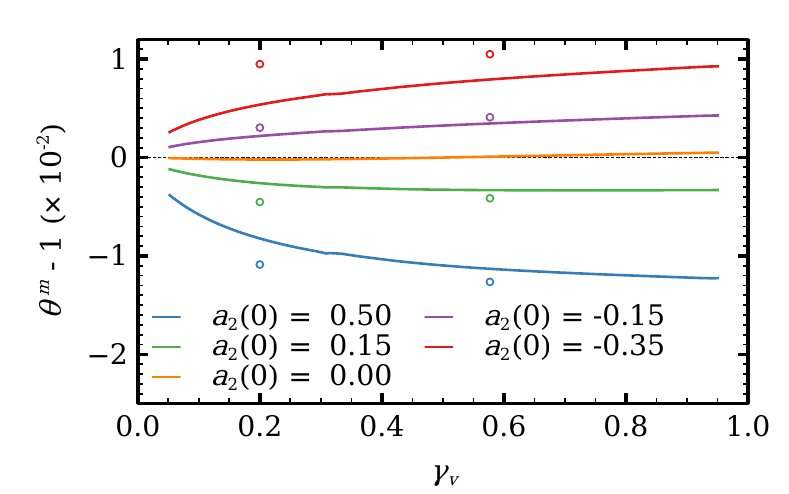}\\
(b)\\
\includegraphics[width=0.48\textwidth]{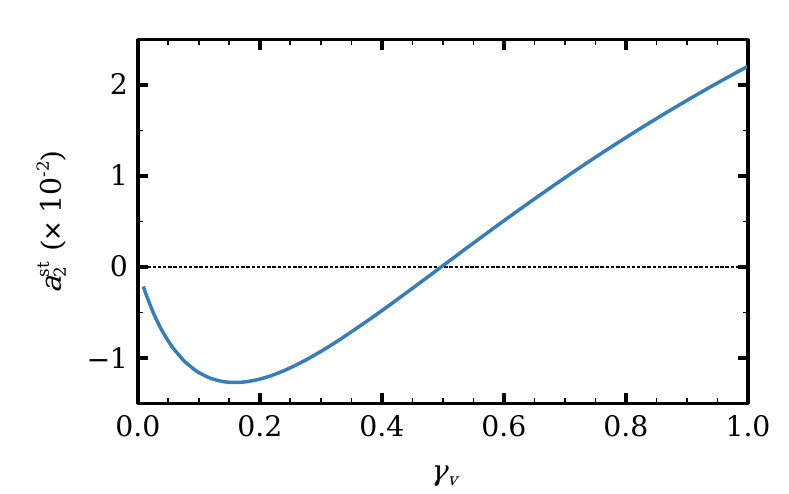}\\
\end{tabular}
\caption{
(a) Size of the theoretical (lines) and MD simulation (open circles) hump in the granular 
temperature evolution as a function of $\gamma_\nu$ for different values of $a_2(0)$ and 
$a_3(0)$ (as reported in Table~\ref{table:kovacs}). The sign of the kurtosis difference 
$a_2(0)-a_2^\mathrm{st}$ determines whether the evolution curves exhibit an upwards 
($a_2(0)<a_2^\mathrm{st}$) or downwards ($a_2(0)>a_2^\mathrm{st}$) hump. (b) The theoretical 
stationary kurtosis $a_2^\mathrm{st}$ is a non-constant function of the inelasticity 
parameter $\gamma_\nu$.} 
\label{fig:hump_size}
\end{figure}

\begin{table}
\caption{\label{table:kovacs} Initial values of the cumulants $a_2$ and 
$a_3$ used in Figure~\ref{fig:kovacs}. The stationary values of the cumulants 
are $a_2^\mathrm{st}=-0.012312$ and $a_3^\mathrm{st}=-0.003249$ for 
$\gamma_\nu=0.2 $ and $a_2^\mathrm{st}=0.003957$ and $a_3^\mathrm{st}=-0.002768$ 
for $\gamma_\nu=0.577$.}
\begin{tabular}{lllll}
\hline\hline
    $a_2(0)$ &   ~0.5 & ~0.15 & $-0.15$ &  $-0.35$\\
    \hline
    $a_3(0)$ & $-0.07143$ & ~0.05357 & $-0.11786$ & $-0.375$\\
    \hline \hline
\end{tabular}
\end{table}

We illustrate this in Figure~\ref{fig:kovacs}, in panel (a) for low inelasticity 
($\gamma_\nu=0.2$) and in (b) for a more inelastic gas ($\gamma_\nu=0.577$); 
the different values of $a_2(0)$ and $a_3(0)$ used in these curves are summarized 
in Table~\ref{table:kovacs}. 
For the two values of $\gamma_\nu$ used here, strong and remarkably similar 
KEs are observed, as
Figure~\ref{fig:kovacs} shows. Note that the hump sign for the curves present
matches that of the difference $a_2^\mathrm{st}-a_2(0)$; i.e., the hump is
downwards for $a_2^\mathrm{st}-a_2(0)<0$ and upwards for $a_2^\mathrm{st}-a_2(0)>0$. 
This behavior is identical in both
dissipative coefficient values analyzed here ($\gamma=0.577$, which has high
dissipation and $\gamma_v=0.2$, with small dissipation). Therefore, it does
not appear to be essentially determined by inelasticity. This is much in contrast with the
behavior for hard spheres, where  a clear hump sign transition for the coefficient of
restitution critical value $\alpha_c=1/\sqrt{2}\sim 0.7$  was reported in the
bibliography \cite{PT14}. It is also interesting to notice that, in general, there is
a good agreement between theory and simulation, so it is guaranteed that the result observed here
is not an artifact of the approximations used in the theoretical approach.

In order to further clarify this behavior, we analyze in
Figure~\ref{fig:hump_size}(a) the evolution of the signed hump size (here represented
as $\theta^m-1$ where $\theta^m\equiv T^m/T^\mathrm{st}$ and $T^m$ is the maximum/minimum value achieved by the
granular temperature during the transient), for different
initial values of the cumulants $a_2(0), a_3(0)$, vs. the dissipative coefficient (as
summarized in Table~\ref{table:kovacs}). Results show that the hump sign does not
depend essentially on dissipation upon collision. Notice that each curve remains with
the same hump sign whether dissipation is high or low. This corroborates that the behavior
of this memory effect for viscoelastic particles and, thus, very possibly in
experiments, is essentially different from that in the less realistic hard sphere
collisional model.

However, a subtle sign transition for the $a_2(0)=0$ curve can be observed 
at $\gamma_\nu\sim0.45$  (orange line in
Figure~\ref{fig:hump_size}(a)). We plot in Figure~\ref{fig:hump_size}(b),  the
difference $a_2^\mathrm{st}-a_2(0)$ for the case $a_2(0)=0$
(i.e., $a_2^\mathrm{st}-a_2(0)=a_2^\mathrm{st}$). We note that, in effect,
$a_2^\mathrm{st}$ changes sign at $\gamma_\nu\sim0.45$. Moreover,
Figure~\ref{fig:hump_size}(b) shows that the sign of
this difference matches the hump sign for all $\gamma_\nu$, by comparison 
with the corresponding curve in
Figure~\ref{fig:hump_size}(a), and therefore, 
that the hump sign is determined by the difference $a_2^\mathrm{st}-a_2(0)$. 
But, even more interestingly, we observe that curve $a_2(0)$  
displays a Kovacs sign transition that is the opposite to the one reported for hard
spheres; i.e., in a granular gas of viscoelastic particles the Kovacs hump is in the
quasi-elastic collision limit.

\begin{figure}[!ht]
\centering 
\includegraphics[width=0.5\textwidth, trim=1mm 0 0 0, clip]{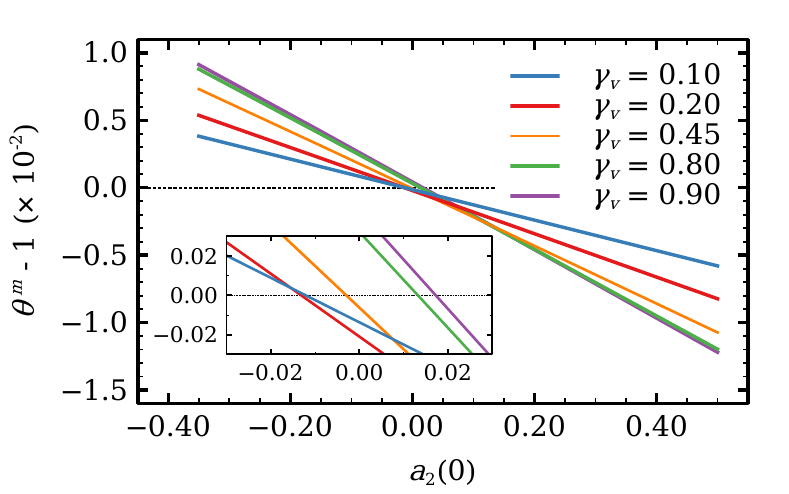}
\caption{The signed hump size ($\theta^m-1$) vs. $a_2(0)$ for different
  dissipative coefficient values ($\gamma_\nu$  in this case).} 
\label{fig:hump_size_2}
\end{figure}

Since, due to the scale in Figure~\ref{fig:hump_size} this detail
is hard to grasp, we plot in Figure~\ref{fig:hump_size_2} the hump size for varying
$a_2(0)$  and several dissipative coefficient ($\gamma_\nu$) values. The large panel
helps to visualize that, regarding the hump sign, the initial value of $a_2(0)$
is indeed by far the most determining factor. Remarkably, it helps to figure also
that it is not the only factor, since the
hump changes sign at slightly different values of $a_2(0)$  for different values of 
the dissipative coefficients (see inset). This feature is also exclusive of our 
system and does not appear in granular
gases of hard spheres, for which the transition is fixed at
coefficient of restitution $\alpha_c=1/\sqrt{2}$ for all thermal protocols; i.e., all
initial state cases (see Figure 9 in \cite{TP14}, for
instance). Figure~\ref{fig:hump_size_2} inset helps to visualize also
that the hump sign, for $a_2(0)\sim0$, changes from negative for quasi-elastic collisions
(smaller $ \gamma_\nu$ values) to positive sign for highly inelastic collisions
(larger $\gamma_\nu$ values). This is evident since, at $a_2(0)=0$, the
curves for more elastic collisions ($\gamma_\nu=0.1; \gamma_\nu=0.2$) have a negative 
hump whereas the curves for more inelastic gases ($\gamma_\nu=0.8; \gamma_\nu=0.9$) 
have a positive hump. 

Moreover, this is a striking result since it implies that the most dissipative granular
gases of viscoelastic particles, and not the quasi-elastic ones, display a thermal 
memory more similar to that of equilibrium systems.

\section{Discussion} \label{discussion}

We have analysed the presence of two memory effects in a system of identical
viscoelastic spherical grains, where the coefficient of restitution depends on the
relative velocities of pairs of colliding particles \cite{DBPB13}. We subject the granular gas to thermal
protocols by means of a homogeneous stochastic force or thermostat. 
We have observed that, under appropriate conditions,
the temperature relaxation curves may display a rich phenomenology of memory effects;
i.e., both the ME and the KE are observed in granular gases of
viscoelastic particles,  like in their analogs with simpler, less
realistic collisional models. 

Nevertheless, we have reported profound differences in the behavior of both memory
effects with respect to inelastic hard spheres. Namely, the ME is
significantly smaller for viscoelastic particles, and the increase of the effect for
more inelastic particles is much more limited. 

It is interesting to note that the KE behavior that we observed is qualitatively 
different from the one reported for granular gases of smooth hard particles, as described 
in the previous section. Much like in the case of a granular gas of hard particles, 
we observe here downwards and upwards KEs, but the hump sign is mainly 
--but not uniquely-- controlled here by the initial values of the cumulants, whereas 
for smooth hard particles the sign transition is driven by
inelasticity \cite{PT14}. Moreover, the hump sign is always positive for more
inelastic collisions for all initial values of the cumulants, except in the region
$a_2(0)\sim0$. Indeed, a hump sign
transition with respect to the dissipative coefficient has been detected for $a_2(0)\sim0$,
from positive hump --for more inelastic particles-- to negative hump --for more elastic
particles--. Strikingly as well, the critical value of the dissipative coefficient for
hump sign transition slightly varies for different initial conditions, contrary to the
case of hard particles, where the transition has been reported to occur at the fixed
value of coefficient of restitution $\alpha_c=1/\sqrt{2}$ for all initial conditions.


\section{Conclusion} \label{conclusion}

We have shown that the ME and/or KE appear when a system is relaxing towards a 
stationary state and the initial values of the cumulants are far from their steady state values. 
We also demostrate numerically that this is due to the existing coupling 
between cumulants and temperature, which induces its direction. 
Furthermore, both memory effects arising from a coupling of the granular temperature 
with $a_2$ and $a_3$ represents a step towards a unifying description of memory effects 
in the context of granular dynamics and opens new avenues for the study of more 
sophisticated systems like the ones composed by anisotropic particles \cite{GBB16,MVC18,LTL19}.

In summary, thermal memory  effects are intrinsically present in a
granular gas whose particle collisions are described with the more realistic 
viscoelastic model, but with important differences with respect to simpler
collisional models. In fact, this is another proof that the ME is not a 
theoretical artifact depending on a particular model. Therefore, and since 
these memory effects are ubiquitous in granular fluid theoretical studies, 
we suggest they should be detectable in
laboratory experiments but also with important differences with  results
reported in previous theoretical developments.

The present work is thus a step towards a more realistic description of memory
effects in granular dynamics, to be compared with the experimental behavior in a future work, in the same train of thought that both effects have been found in single particle experiments.
Insight into thermal relaxation will improve the understanding of the granular dynamics, whose transport properties are known to be highly dependent on temperature. Since grains are a component present in a variety of industrial processes \cite{G03,BP04}, applied research is expected to benefit from these results as well.
\bigskip

\begin{acknowledgments}
This work has been partially funded by the Spanish Ministerio de Ciencia, Innovaci\'on y 
Universidades and the Agencia Estatal de Investigaci\'on through grants No. 
MTM2017-84446-C2-2-R (A.L., E.M. and A.T.), FIS2017-84440-C2-2-P (A.T.), and 
FIS2016-76359-P. (F.V.R.). M.L.-C. and F.V.R. also acknowledge support from the 
regional Extremadura Government through projects No. GR18079 \& IB16087. Computing 
facilities from Extremadura Research Centre for Advanced Technologies (CETA-CIEMAT) 
are also acknowledged. All grants and facilities were provided with partial support 
from the ERDF. 
\end{acknowledgments}

\section*{Data Availability Statement}

The data that support the findings of this study are available from the corresponding author upon reasonable request.


\appendix*

\renewcommand{\theequation}{A.\arabic{equation}}

\section{Numerical methods}

As explained in the body of the text, we have carried out our study by using 
three complementary techniques: numerical solution (by means of the MATLAB package) 
of the time differential equation system of the first three moments of the distribution 
function, accounting for the required approximations, molecular dynamics (MD) 
simulations \cite{PS05}, and numerical solution of the kinetic equation obtained 
through the Direct Simulation Monte Carlo  method (DSMC) \cite{B94}.

Solutions to equations~\eqref{eq:main_ode1}-\eqref{eq:main_ode3} were numerically approximated using a forward Euler method with a time step $\Delta \tau = 0.001$. The method's expected first order of convergence was numerically verified as follows: let $u$ denote the solution to one of the differential equations, and let $u_{h}$ be the numerical approximation to $u$ using time steps of length $h$. As the actual solution is not known beforehand, for any length $h$ we shall store the approximated values at times $\tau = n\Delta\tau$, for $n=0,1,2,\ldots$, into a vector $v_h$. That is, $v_h^n \approx u(n\Delta\tau)$. If the method has an order of convergence $p$, then $\max_{n}|u(n\Delta\tau)-v_h^n|\leq C h^p$, for some constant $C>0$. Since $u$ is not available, the order can be estimated by means of successive refinements of the time step length:
\begin{equation}
    \frac{\|v_{\Delta\tau}-v_{\Delta\tau/2}\|_{\infty}}{\|v_{\Delta\tau/2}-v_{\Delta\tau/4}\|_{\infty}} = 2^p + O( \Delta\tau),
\end{equation}
where $\|v\|_\infty=\max_n|v^n|$. Using this, the order of convergence $p\approx 1$ has been consistently recovered for all the experiments described heretofore. Moreover, a time step $\Delta\tau \leq 0.01$ was sufficient in all cases. Hence $\Delta\tau = 0.001$ lies well within the stability region of the method.

MD computer simulation is a powerful tool that allows knowing the 
spatial coordinates of every constituent particle in the molecular gas at all 
times, an option that, though time-consuming, is barely accessible through 
experiments. It numerically solves Newton's equations of motion for all particles
by accounting for forces and torques acting on them and by incorporating realistic
values for the parameters describing the material properties.

In (event-driven) MD simulations, collisions are binary and of infinitesimal 
duration, and the dynamics is regulated by a sequence of discrete events. 
The system particles follow known ballistic trajectories during the time intervals
between collisions, allowing the computation of particle positions on the next 
collision in a single step.  The MD algorithm can be summarized as:
\begin{enumerate}
    \item Initialization of positions and velocities for all $N$ particles 
    at time $t=0$.
    \item Determination of the time $t_{next}$ at which the next collision 
    occurs and the particles $i$ and $j$ involved in such a collision.
    \item Update of positions of all particles at time $t_{next}$.
    \item Update of velocities of the two colliding particles, according to 
    the system collision rule, 
     in our case given by:
    \begin{eqnarray}
      v_i'=v_i - \dfrac{1+\epsilon}{2}\left(v_{ij} \cdot 
        e_{ij} \right) e_{ij}, \\ v_j'=v_j - \dfrac{1+\epsilon}{2}\left(v_{ij} \cdot 
        e_{ij} \right) e_{ij}.
    \end{eqnarray}
    Here, the not-primed and primed $v_k$s refer to pre- and 
    post-collisional velocities, respectively, $\epsilon$ 
    is the restitution coefficient given in equation 
    \eqref{restitution}, $v_{ij}= v_{i} - v_{j}$ 
    is the relative 
    velocity and $e_{ij}$ is a unit vector pointing from the center of mass 
    of particle $j$ to that of particle $i$ at the instant of the collision.
\end{enumerate}
Steps (ii)--(iv) are repeated until a steady state, which is defined by a target 
temperature provided by a thermostat, is reached.

In our experiments, the MD data sets have been obtained out of simulation runs 
for systems with dimensionless number density
$n\sigma^3= 0.01$, with  $n$ the number of particles per volume unit, which are
composed of particles of mass and diameter equal to 1. 
We have considered periodic boundary conditions to mimic an 
infinitely extended system.
In the ME study, we have used a coefficient of 
restitution with dissipative coefficient $\gamma_\nu=0.577$ and an 
appropriate gamma density function for the initial velocity distribution  
\cite{HC78} to recreate the selected values of $a_2$.
Hence, the values of $a_3$ are the ones associated to such a distribution. 
For ME simulations, we have averaged over $10^3$
trajectories. On the other hand,  more computational trajectories are needed 
to evince the more subtle KE for some of the values of $\gamma_\nu$, thus we 
have averaged over $10^4$ trajectories for each of the reported dissipative 
coefficient values, $\gamma_\nu=0.20$ and $0.577$.

As a brief description of the DSMC method, one must remember that the dynamics 
of the system is determined by the Boltzmann equation:
\begin{equation} \label{eq:boltzmann}
    \left[ \partial_{t} + (v \cdot \grad ) \right] f + \frac{1}{m} 
    \frac{\partial}{\partial v} \cdot (F f)= J[f,f].
\end{equation}

 To solve this equation, rather than exactly calculating the outcome of every 
 impact, the Monte Carlo approach generates collisions stochastically. These 
 are not necessarily real collisions but outcome makes the system behaviour 
 be correct around mean free path length scales \cite{Alexander1997}.

 The standard algorithm (using Bird's no-time-counter, NTC, scheme \cite{Bird1994}) consists of the following steps: 

\begin{enumerate}
	\item \textbf{Initialization}: As with MD, $N$ particles 
	are randomly created and distributed (with positions $r_i$) among the simulation volume. 
	One key difference though is that there is no need to care for overlapping 
	particles since the computed collisions will not be physically exact. A 
	large number of particles can be simulated in DSMC, typically from $10^4$ 
	to $10^8$. However, if the number of grains is fewer than about $20$ per 
	cubic mean free path the results may not be accurate ($\lambda =(\sqrt 2 
	\pi \sigma^2 n)^{-1}$   in a dilute gas). Besides positions, each particle 
	is given an initial velocity $v_i$, usually extracted from a 
	Maxwellian distribution function.
	\item \textbf{Collision}: A random number of particles are selected for 
	colliding; this selection method is derived from kinetic theory. For 
	inhomogeneus systems, particles very distant from each other should not 
	interact, so the simulation box should be divided into cells and collisions 
	are evaluated only among particles of the same sub-cell (this is not our case, 
	there are no long-range gradients in our system that require grid sub-division). 
	This set of random representative collisions is processed at each time 
	step $\Delta t$ (which is set at a value much smaller than the mean free time); 
	only the magnitude of the relative velocity between 
	particles is used to evaluate the probability of a collision, regardless 
	of their positions. Therefore, the probability of a collision between a 
	pair of particles is:
	\begin{equation} \label{eq:dsmc_coll_prob}
	    P_{col}(i,j) = \frac{\abs{v_{ij}}}{\sum_{m=1}^{N_c}
	    \sum_{n=1}^{m-1} 
	    \abs{v_{mn}}},
	\end{equation}
	where $N_c$ is the number of particles in that sub-cell.
	
	Since using previous equation is very inneficient due to the iteration 
	over every particle in the denominator, another method is used for selecting 
	which collisions to evaluate.
	
	First, a random uniform value $r\in(0,1)$ is generated, then an aleatory 
	pair of particles $i,j$ is selected. That pair is considered to undergo a collision if the following condition is met:
	\begin{align}
	\frac{\abs{v_{ij}}}{v_{\mathrm{r,max}}} > r ,
	\end{align}
	where $v_{\mathrm{r,max}}$ is the maximum relative speed in the cell. If 
	the pair $i,j$ is accepted, post-collisional velocities are computed for 
	that pair. Finally, this algorithm is repeated until a certain number 
	$M_{col}$ of collisions has been accepted and processed. The value of 
	$M_{col}$ is costly to calculate, thus, the algorithm should be repeated 
	a number of times given by $M_{cand}$, for a hard-sphere model:
	\begin{align}
	    M_{cand} = \frac{N_{c}^2 \pi \sigma^2 v_{\mathrm{r,max}} \Delta t}{2 V_c},
	\end{align}
	where $V_c$ is the volume of the cell. Special attention must be paid to 
	the estimation of $v_{\mathrm{r,max}}$,  which is very computationally expensive, 
	so usually $v_{\mathrm{r,max}}$ is indirectly overestimated by multiplying 
	$\langle v \rangle$ by a certain factor.
	
\end{enumerate}

 These steps are repeated until the system reaches a steady state; to achieve 
 this, a thermostat has been implemented following the methods described in 
 \cite{Montanero2000}. Finally, the viscoelastic behaviour of our particles has 
 been encoded in the operator $J[f,f]$ which defines the outcome of collisions. 
 In this paper, for DSMC simulations we have used 100 statistically independent 
 replicas  with $N=2\times 10^5$ particles each, a number which has proven to 
 yield small errors in other works \cite{Maltsev2011}.


%
%

%


\bibliography{aipsamp}

\begin{thebibliography}{70}
\bibitem [{a3p()}]{a3plays}%
  \BibitemOpen
  \href@noop {} {}\bibinfo {note} {Notice that, contrary to the case of
  inelastic smooth hard particles [ref. 31] the cumulant $a_3$ for viscoelastic
  particles is in general not negligible in comparison with $a_2$ [ref. 17] and
  for this reason we consider both cumulants in the relaxation
  process.}\BibitemShut {Stop}%
\end{thebibliography}

\end{document}